\documentstyle[epsfig,aps,prl,twocolumn]{revtex}
\newcommand{\etal}{{\em{et al.}}}
\newcommand{\Vud}{V_{ud}}
\newcommand{\Vus}{V_{us}}
\newcommand{\Vub}{V_{ub}}
\newcommand{\Vcd}{V_{cd}}
\newcommand{\Vcs}{V_{cs}}
\newcommand{\Vcb}{V_{cb}}
\newcommand{\Vtd}{V_{td}}
\newcommand{\Vts}{V_{ts}}
\newcommand{\Vtb}{V_{tb}}
\newcommand{\VTD}{|V_{td}|}
\newcommand{\VTS}{|V_{ts}|}
\newcommand{\VTB}{|V_{tb}|}
\newcommand{\forcebigbraces}{\phantom{\rule[-0.5em]{0.001pt}{2.2em}}} 
\newcommand{\mb}{m_b}
\newcommand{\mt}{m_t}
\newcommand{\mH}{m_H}
\newcommand{\mZ}{m_Z}
\newcommand{\mW}{m_W}
\begin{document}
%
%
%
\draft
\title{First determination of the quark mixing matrix element ${\mathbf{\Vtb}}$\\
       independent of assumptions of unitarity}
%
%
\author{John Swain and Lucas Taylor}
\address{Department of Physics, Northeastern University, Boston, Massachusetts 02115, USA}
\date{\today}
\maketitle
\begin{abstract}
We present a new method for the determination of the Cabibbo-Kobayashi-Maskawa 
quark mixing matrix element $\VTB$ from electroweak loop corrections, 
in particular those affecting the process $Z\rightarrow b\bar{b}$.
From a combined analysis of results from the LEP, SLC, Tevatron, 
and neutrino scattering experiments we determine 
$\VTB = 0.77^{+0.18}_{-0.24}$.
This is the first determination of $\VTB$
which is independent of unitarity assumptions.
\end{abstract}
%
\pacs{%
12.15.Ff,          
12.15.Lk,          
13.38.Dg,          
14.65.Ha}          
%
%
%
\section{Introduction}
The Cabibbo-Kobayashi-Maskawa (CKM) matrix\cite{CKM} describes the relationship between
weak and mass eigenstates of quarks, assuming that there are three generations. 
By convention, up-type quarks are unmixed such that all the mixing is expressible
in terms of the $3\times 3$ unitary matrix $V$ 
\begin{equation}
  \left( \begin{array}{c} 
           d' \\
           s' \\
           b'
         \end{array}
  \right)  =
  \left( \begin{array}{ccc}
      \Vud & \Vus  &  \Vub  \\
      \Vcd & \Vcs  &  \Vcb  \\
      \Vtd & \Vts  &  \Vtb  
   \end{array}  \right)  
  \left( \begin{array}{c} 
           d \\
           s \\
           b
        \end{array} \right),
  \label{eq:ckm}
\end{equation}
where unprimed states denote mass eigenstates and primed ones
denote weak eigenstates.
The unitarity of the CKM matrix is an assumption which must be 
subjected to experimental verification by independent measurements 
of the elements.
Non-unitarity of the matrix would be a clear signature for physics 
beyond the standard model, such as a fourth generation or 
non-universality of the quark couplings.
Of particular interest in terms of possible new physics
is the element $\Vtb$ which describes the coupling of the 
two heaviest quarks.

It is often assumed that $\VTB \approx 1$ although this element has never 
been determined without assumptions of unitarity.
Assuming that there are only three generations and 
unitarity of the CKM matrix yields:
$0.9989 < \VTB < 0.9993$ at the 90\% confidence level~\cite{PDG96SHORT}. 
Relaxing the assumption of three generations but maintaining that
of unitarity yields $0 < \VTB < 0.9993$ at the 90\% confidence level~\cite{PDG96SHORT},
while relaxing also that of unitarity leaves $\VTB$ unbounded.

In this paper we describe a new method for the determination of $\VTB$
from electroweak loop corrections, in particular to the 
process $Z\rightarrow b\bar{b}$.
From a combined analysis of data from the LEP, SLC, Tevatron,
and neutrino scattering experiments,
we determine for the first time the value of $\VTB$, without any assumptions 
about the unitarity of the CKM matrix.
The implications for other quantities, such as the the top mass, Higgs 
mass, and the strong coupling constant, are discussed.

\section{Effects of ${\mathbf{\Vtb}}$ on electroweak radiative corrections}
\begin{figure}[!tbp]
\begin{center}
    \mbox{\epsfig{file=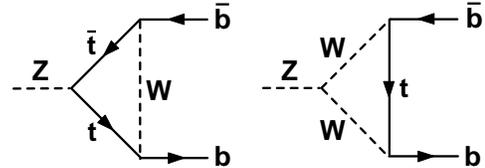,width=0.35\textwidth,clip=}}
\end{center}
  \caption{Vertex correction diagrams, at order one-loop, which contribute to 
           the partial width for $Z\rightarrow b\bar{b}$.}\label{fig:feynman_diags}
\end{figure}

The precision of the electroweak data from the LEP and SLC experiments
is sufficient to be sensitive to weak loop diagrams involving the top quark. 
The top quark appears in $Z$ vacuum polarisation loops, 
thereby affecting all the $Z$ partial widths,
and in the GIM-suppressed vertex diagrams shown at the one-loop level
in Fig.~\ref{fig:feynman_diags} 
which affect the $Z$ partial width, $\Gamma_{b\bar{b}}$,
for the process $Z\rightarrow b\bar{b}$.
From fits to the $Z$ electroweak parameters, the top quark mass has been determined to be 
$\mt = 158^{+14}_{-11}$\,GeV~\cite{WARD_EPS97}, 
in agreement with the direct measurement from the Tevatron of 
$\mt = 175.6 \pm 5.5$\,GeV~\cite{RAJA}.

Hitherto the theoretical treatments of weak loop corrections to 
$Z$ decay processes have assumed that $\VTB = 1$; we relax this assumption.
Following the treatment of Barbieri, Beccaria, Ciafaloni, Curci, and 
Vicer{\'{e}} (BBCCV)~\cite{BARBIERI93ALL}, $\Gamma_{b\bar{b}}$ may be written as:
\begin{eqnarray}
\Gamma_{b\bar{b}} & = & 
\frac {G_\mu \mZ^3} {8\pi{\sqrt{2}}} \rho R_{\mathrm{QED}} R_{\mathrm{QCD}}
           {\sqrt{1 - \frac{4 \mb^2}{\mZ^2}}} \nonumber \\
       & & \times \left[ 
                        \left( g^2_{bV} + g^2_{bA} \right)
                        \left( 1 + 2 \frac{\mb^2}{\mZ^2}             \right)
                               - 6 g^2_{bA} \frac{\mb^2}{\mZ^2}
                  \right],
\end{eqnarray}
where 
$G_\mu$ is the Fermi constant,
$m_Z$ and $m_b$ are the masses of the $Z$ and the $b$-quark 
respectively, and $\rho$ includes the effects of radiative 
corrections to the $Z$ propagator.
$R_{\mathrm{QED}}$ and $R_{\mathrm{QCD}}$, which are approximately unity, 
describe the QED and QCD vertex corrections~\cite{PREC_CALC,ZFITTER}.
The couplings, $g_{bV}$ and $g_{bA}$ incorporate vertex 
corrections described by the parameter, $\tau$, as follows:
\begin{eqnarray}
g_{bA}           & = & 1 + \tau                                  \\
g_{bV}           & = & 1 - \frac {4}{3} s^2 + \tau               \\
{\mathrm{where:~~~}}  s^2 & = & \frac{1}{2} 
                             \left( 1 - {\sqrt{ 1 - \frac {4 \pi \alpha} {{\sqrt{2}} G_\mu \mZ^2 \rho}}}
                             \right).  
\end{eqnarray}
We have taken the results of the BBCCV calculations up to two-loops
and included the effects of $\Vtb$ at the one-loop level, such that $\rho$ 
is unchanged while $\tau$ is modified by a multiplicative factor of $\VTB^2$.
In the limit $\mt \gg \mH $, $\tau$ is given by:
\begin{eqnarray}
\tau          & = & -2 \VTB^2 x \left[1 +\frac{x}{3}\left(27 - \pi^2\right)\right], 
\label{eqn:barbtau1}
\end{eqnarray}
where $x = {G_\mu \mt ^2}/{8\pi^2\sqrt{2}}$.
In the limit $\mt \ll \mH $, $\tau$ is given by: 
\begin{eqnarray}
\tau      & = &             -2 \VTB^2 x                                                      \nonumber \\
          &               & \times 
                            \left\{1 + \frac{x}{144} \left[
                            \forcebigbraces
                            \right. \right.                                                    
                             311 + 24 \pi^2 + 282 \log{r} + 90 \log^2{r}  
                                                                                             \nonumber \\
          &               &  - 4 r 
                            \left(  40 + 6 \pi^2 + 15 \log{r} + 18\log^2{r}
                            \right)                                                          \nonumber \\
          &               & + \frac{3 r^2}{100} 
                            \left(24209 - 6000 \pi^2 -
                            45420 \log{r}                                                   
                            \right.                                                          \nonumber \\
          &               & \left.\left.\left.                            
                            - 18000 \log^2{r}
                            \right)
                            \forcebigbraces
                            \right]\right\},                                       
\label{eqn:barbtau2}
\end{eqnarray}
where $r={\mt ^2}/{\mH ^2}$.
In the intermediate region ($\mt \approx \mH$) 
$\tau$ is described by a polynomial parametrisation
of the full BBCCV calculation, as a function of $\mt/\mH$, 
multiplied by a factor of $\VTB^2$.
At the two-loop level some diagrams are of the order $\VTB^4$.
Our treatment of these to order $\VTB^2$ is justified by their relatively small contribution
and the current sensitivity of the experimental data, as will be seen below.

\section{Determination of ${\mathbf{\VTB}}$ from a fit to electroweak data}
The BBCCV corrections are incorporated in the ZFITTER 
program~\cite{PREC_CALC,ZFITTER} which is used by the 
LEP/SLC electroweak working groups to derive parameters 
such as $\mt$ and $\mH$ from $Z$ data~\cite{WARD_EPS97}.
We make modest modifications to ZFITTER to allow for 
the effects of $\Vtb$ described in the previous section.

The $Z$ parameters, from a combined fit to LEP/SLC data, 
which we use as input are~\cite{WARD_EPS97}: 
the $Z$ mass,                                         $\mZ$;
the $Z$ width,                                        $\Gamma_Z$;
the hadronic pole cross-section,                    $\sigma^0_h$;
$R_\ell \equiv \Gamma_{\mathrm{had}} / \Gamma_{\ell\ell}$
where $\Gamma_{\mathrm{had}}$ is the hadronic partial 
$Z$ width and $\Gamma_{\ell\ell}$ is the leptonic 
partial width, assuming lepton universality;
and the leptonic pole forward-backward charge
asymmetry assuming lepton universality, $A^{0,\ell}_{\mathrm{FB}}$.
The parameter values, their errors, and their correlation coefficients
are shown in Table~\ref{tab:lineshape}.

The parameters pertaining to $b$ and $c$ quarks which we 
use are~\cite{WARD_EPS97}:
$R^0_b \equiv \Gamma_{b\bar{b}} / \Gamma_{\mathrm{had}}$;                                    
$R^0_c \equiv \Gamma_{c\bar{c}} / \Gamma_{\mathrm{had}}$;                 
$A^{0,b}_{\mathrm{FB}}$ and 
$A^{0,c}_{\mathrm{FB}}$, 
the forward-backward charge asymmetries at the $Z$ pole
for $b$ and $c$ quarks respectively;
and $A_f$ $(f=b,c)$ where 
$A_f \equiv 2 g_{fV} g_{fA} /
(g^2_{fV} + g^2_{fA})$.
The parameter values, their errors, and their correlation coefficients
are shown in Table~\ref{tab:bcpars}.

We also use the following parameters which are to a good 
approximation experimentally uncorrelated:
$A_\tau \equiv -P_\tau$, the average tau polarisation~\cite{WARD_EPS97};
$A_e$ from the tau polarisation forward-backward asymmetry~\cite{WARD_EPS97}; 
$A_{\mathrm{LR}}$, the left-right asymmetry from SLD~\cite{WARD_EPS97};
the QED coupling constant, $\alpha(\mZ)$~\cite{EIDELMANN95A};
the strong coupling constant $\alpha_{\mathrm{s}}(\mZ)$~\cite{PDG96SHORT}
where the value obtained from the $Z$ width is not included;
the $W$ masses from LEP II~\cite{WARD_EPS97}, 
                  CDF~\cite{WMASSCDF97}, 
                  D\O~\cite{WMASSDZERO97}, and 
                  UA2~\cite{WMASSUATWO}, 
averaged according to Ref.~\cite{PIC97};
the top mass from the Tevatron~\cite{RAJA};
and $(1-\mW^2/\mZ^2)$ from 
$\nu$N scattering measurements by CHARM~\cite{CHARM}, 
                                  CDHS~\cite{CDHS}, and 
                                  CCFR~\cite{CCFR}.
The parameter values and their errors
are shown in Table~\ref{tab:otherpars}.

For given values of $\mZ$, $\mt$, $\mH$, $\alpha_{\mathrm{s}}$, $\alpha$, and $\VTB$
our modified version of ZFITTER provides predictions for all of the 
parameters shown in Tables~\ref{tab:lineshape},~\ref{tab:bcpars}, and~\ref{tab:otherpars}.
These predictions and the corresponding measured quantities, together with 
their associated errors and correlation coefficients,
are used to construct a chisquare probability  
$\chi^2 (\mZ,\mt,\mH,\alpha_{\mathrm{s}},\alpha,\Vtb)$.
The minimum of the chisquare is then determined numerically.
As a technical cross-check, we use the same input parameters as 
those of Ward~\cite{WARD_EPS97}, set $\VTB=1$, and successfully reproduce the  
results for $\mZ$, $\mt$, $\mH$, $\alpha_{\mathrm{s}}$, and $\alpha$.

The results of the fit with $\VTB$ free are shown in Table~\ref{tab:results}.
We determine $\VTB = 0.77^{+0.18}_{-0.24}$.
This value is consistent with the unitarity prediction
of $\VTB \approx 0.9991$ to within approximately one 
standard deviation. 
For comparison, Table~\ref{tab:results} also includes 
the results with $\VTB$ fixed to $0.9991$.
In both fits, the $\chi^2$ probability is consistent with 
expectations given the number of degrees of freedom.

\section{Discussion}
\begin{figure}[!tbp]
\noindent\mbox{\epsfig{file=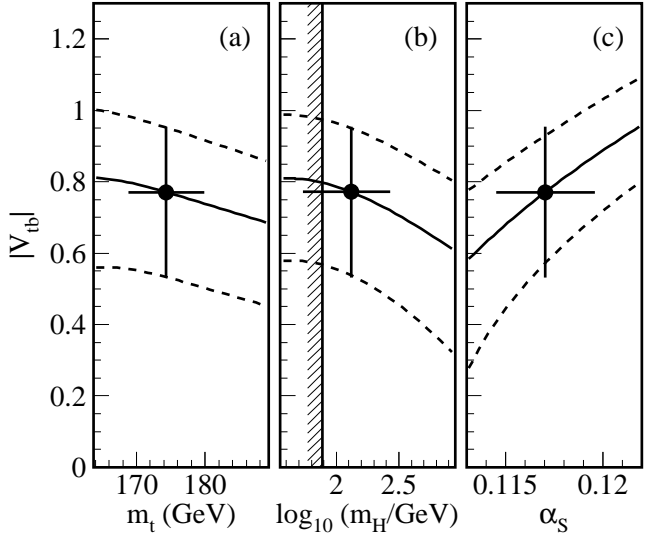,width=0.48\textwidth,clip=}}
  \caption{Variation of $\VTB$ with 
           (a) $\mt$, 
           (b) $\log_{10}(\mH/{\mathrm{GeV}})$, and 
           (c) $\alpha_{\mathrm{s}}$.
           The point with error bars denotes the result of the 
           fit allowing for all errors and correlations.
           The solid line shows the dependence of $\VTB$ on the 
           ordinate variable; the dashed lines correspond to the 
           68\% confidence level.
           The hatched line of (b) shows the low $m_H$ region
           excluded by direct searches at LEP II.}\label{fig:vtb_fit}
\end{figure}

The fitted values of $\mZ$ and $\alpha(\mZ)^{-1}$
are insensitive to $\VTB$ as expected, as shown by the correlation
coefficients from the fit with $\VTB$ free in Table~\ref{tab:correl}.   
The anti-correlations of $\VTB$ with $\mt$ and $\mH$,
shown in Fig.~\ref{fig:vtb_fit}(a) and \ref{fig:vtb_fit}(b) and 
in Table~\ref{tab:correl}, 
have only a weak effect on the determinations of $\mt$ and $\mH$ since 
the Tevatron measurement of $\mt$ and the vacuum 
polarisation contribution to the $Z$ width constrain 
$\mt$ and $\mH$ independent of $\VTB$.
Nonetheless, allowing $\VTB$ to float increases the fitted 
values of $\mt$ by 1.5\,GeV and of $\mH$ by approximately 30\,GeV.
Allowing $\VTB$ to float decreases the fitted value of 
$\alpha_{\mathrm{s}}$ by approximately 0.7 standard deviations,
due to the fairly strong correlation of these two quantities,
as shown in Fig.~\ref{fig:vtb_fit}(c) and in Table~\ref{tab:correl}.

To assess the future sensitivity of this technique for determining  
$\VTB$ we reduce the error by a factor of two 
on each of the input parameters in turn, without changing the 
errors on the other parameters.
The only parameters which cause $\Delta\VTB/\VTB$ to change by a relative amount of 
more than 10\%, from the original value of $\Delta\VTB/\VTB\approx 26.5\%$, are 
$\Gamma_Z$,
$R^0_b$, and
$\alpha_{\mathrm{s}}(\mZ)$, which yield uncertainties of
$\Delta\VTB/\VTB \approx$ 22.5\%, 19.9\%, and 22.4\% respectively. 
Factors of somewhat less than two may be expected from 
the final analyses of the LEP I and SLC data samples.
We estimate that the error on $\VTB$ from the final $Z$ samples of 
the LEP and SLC experiments will be approximately 20\%.

Recently CDF presented preliminary results of an analysis of
their $t\bar{t}$ event samples.
However, they necessarily assumed that there are only three 
generations and that $\VTD^2 + \VTS^2 + \VTB^2 =1$,
as required by unitarity,
to extract $\VTB^{3{\mathrm{gen}}} = 0.99 \pm 0.15$~\cite{HEINSON97A}.
Ultimately, the measurement of the single top quark production rate at 
hadron colliders should be sensitive to $\VTB$ without requiring 
such assumptions of unitarity.
The estimated sensitivity at the end of the Tevatron Run II for 
$\delta\VTB/\VTB$ is 12\%--19\%, depending on the uncertainty of the 
gluon structure functions~\cite{HEINSON97B}.


\section{Summary}
We describe a new technique for the determination of the CKM matrix 
element $\VTB$ using loop corrections to electroweak processes, without 
using any unitarity constraints.
From a combined analysis of data from the LEP, SLC, Tevatron and neutrino 
scattering experiments we determine $\VTB = 0.77^{+0.18}_{-0.24}$    
where the error includes the experimental errors and uncertainties on the 
top mass, the Higgs mass, and the strong coupling constant. 
The wider implications of this measurement are discussed 
elsewhere~\cite{CKM_ALL}.

\section*{Acknowledgements}
We would like to thank Joachim Mnich for many valuable discussions,
the National Science Foundation for financial support,
and the Department of Physics, Universidad Nacional de La Plata for 
their generous hospitality while this work was being completed. 




\begin{references}
%
\bibitem{CKM}
N. Cabibbo, Phys. Rev. Lett. {\bf 10},  531  (1963), {M. Kobayashi and T.
  Maskawa, Prog. Theor. Phys. {\bf{49}}, 652 (1973)}.
%
\bibitem{PDG96SHORT}
{R. M. Barnett {\em{et al.}}}, Phys. Rev. {\bf D54},  1  (1996).
%
\bibitem{WARD_EPS97}
D. Ward, {Invited talk at the {\em{International Europhysics Conference on High
  Energy Physics}}, Jerusalem, 1997, (unpublished)}.
%
\bibitem{RAJA}
R. Raja,  in {\em Proceedings of the XXXIInd Rencontres de Moriond, Moriond, 1997} 
edited by J. Tr{\^{a}}n Thanh V{\^{a}}n, 
Editions Fronti{\`{e}}res, Gif-sur-Yvette, (1997).
%
\bibitem{BARBIERI93ALL}
{R. Barbieri \etal, Phys. Lett. {\bf{B288}}, 95 (1992); {\bf B312}(E), 511
  (1993); Nucl. Phys. {\bf{B409}}, 105 (1993)}.
%
\bibitem{PREC_CALC}
D. Bardin {\it et~al.}, in CERN Report {\bf{95-03}}, 
 edited by D. Bardin, W. Hollik and G. Passarino, (1995), (unpublished).
%
\bibitem{ZFITTER}
D. Bardin {\it et~al.}, {CERN-TH {\bf{6443-92}}, (1992), (unpublished);
                        {\bf{hep-ph/9412201}}, (1994), (unpublished)}.
%
\bibitem{EIDELMANN95A}
S. Eidelmann and F. Jegerlehner, Z. Phys. {\bf C67},  585  (1985).
%
\bibitem{WMASSCDF97}
{F. Abe \etal, 
               Phys. Rev. Lett. {\bf{65}},  2243 (1990); 
                                {\bf{75}},    11 (1995); 
               Phys. Rev.       {\bf{D43}}, 2070 (1991); 
                                {\bf{D52}}, 4784 (1995); 
               D. Errede (private communication)}.
%
\bibitem{WMASSDZERO97}
{S. Abachi \etal, Phys. Rev. Lett. {\bf{77}}, 3309 (1996); 
 D. Wood (private communication)}.
%
\bibitem{WMASSUATWO}
{J. Alitti \etal}, Phys. Lett. {\bf B276},  365  (1992).
%
\bibitem{PIC97}
{L. Taylor}, hep-ex/{\bf{9712016}}. 
To appear in the 
{\em{Proceedings of the XVIIth International Conference on
Physics in Collision}} (1997), edited by H. Heath, 
World Scientific, Singapore.
%
\bibitem{CHARM}
{J.V. Allaby \etal, Phys. Lett. {\bf{177}}, 446 (1986); 
                    Z. Phys.    {\bf{C36}}, 611 (1987)}.
%
\bibitem{CDHS}
{H. Abramowicz \etal, Phys. Rev. Lett. {\bf{57}},  298 (1986); 
 A. Blondel \etal,    Z. Phys.         {\bf{C45}}, 361 (1990)}.
%
\bibitem{CCFR}
{K. S. McFarland \etal}, FNAL-Pub-97 {\bf 001-E}, (1997), (unpublished).
%
\bibitem{HEINSON97A}
A.P. Heinson, {In {\em{Proceedings of the 2nd International Conference on B
  Physics and CP Violation, Honolulu, March, 1997}}, {\bf hep-ex/9707026}} (unpublished).
%
\bibitem{HEINSON97B}
{A.P. Heinson, A.S. Belyaev and E.E. Boos}, Phys. Rev. {\bf D56},  3114  (1997).
%
\bibitem{CKM_ALL}
{J. Swain and L. Taylor}, hep-ph/{\bf{97xxxxx}} (unpublished).
%
\end{references}

%
%
%

\clearpage

%
%

%
%
%
\begin{table}[!htbp]
\caption{Measured values and the correlation matrix for 
         $\mZ$, 
         $\Gamma_Z$, 
         $\sigma^0_h$, 
         $R_\ell$, and 
         $A^{0,\ell}_{\mathrm{FB}}$ 
         from a combined fit to LEP and SLC data.}   
\label{tab:lineshape}
{ 
\begin{tabular}{lcrrrrr}                                
\multicolumn{1}{c}{~}                           &
\multicolumn{1}{c}{Measured}                    &
\multicolumn{5}{c}{Correlation coefficient}                \\   \cline{3-7} 
\multicolumn{1}{c}{~}                           &
\multicolumn{1}{c}{value}                       &
\multicolumn{1}{c}{$\mZ$}            &
\multicolumn{1}{c}{$\Gamma_Z$}       &
\multicolumn{1}{c}{$\sigma^0_h$}     &
\multicolumn{1}{c}{$R_\ell$}                    &
\multicolumn{1}{c}{$A^{0,\ell}_{\mathrm{FB}}$}             \\ \hline
$\mZ$                       (GeV) & $ 91.1867 \pm 0.0020 $ & $  1.00 $ & $  0.05 $ & $ -0.01 $ & $ -0.02 $ & $ 0.06 $ \\
$\Gamma_Z$                  (GeV) & $ ~2.4948 \pm 0.0025 $ & $  0.05 $ & $  1.00 $ & $ -0.16 $ & $  0.00 $ & $ 0.00 $ \\
$\sigma^0_h$                (nb)  & $ 41.486~ \pm 0.053~ $ & $ -0.01 $ & $ -0.16 $ & $  1.00 $ & $  0.14 $ & $ 0.00 $ \\
$R_\ell$                          & $ 20.775~ \pm 0.027~ $ & $ -0.02 $ & $  0.00 $ & $  0.14 $ & $  1.00 $ & $ 0.01 $ \\
$A^{0,\ell}_{\mathrm{FB}}$        & $ ~0.0171 \pm 0.0010 $ & $  0.06 $ & $  0.00 $ & $  0.00 $ & $  0.01 $ & $ 1.00 $ \\	  
\end{tabular}
}
\end{table}
%

%
%
%
\begin{table}[!htbp]
{ 
\caption{Measured values and the correlation matrix for 
         $R^0_b$,                
         $R^0_c$,                
         $A^{0,b}_{\mathrm{FB}}$,
         $A^{0,c}_{\mathrm{FB}}$,
         $A_b$, and                  
         $A_c$ 
         from a combined fit to LEP and SLC data.}  
\label{tab:bcpars}
\begin{tabular}{lcrrrrrr}                                
\multicolumn{1}{c}{~}                                      &
\multicolumn{1}{c}{Measured}                               &
\multicolumn{6}{c}{Correlation coefficient}                  \\   \cline{3-8} 
\multicolumn{1}{c}{~}                                      &
\multicolumn{1}{c}{value}                                  &
\multicolumn{1}{c}{$R^0_b$}                     &
\multicolumn{1}{c}{$R^0_c$}                     &
\multicolumn{1}{c}{$A^{0,b}_{\mathrm{FB}}$}     &
\multicolumn{1}{c}{$A^{0,c}_{\mathrm{FB}}$}     &
\multicolumn{1}{c}{$A_b$}                       &
\multicolumn{1}{c}{$A_c$}                          \\   \hline
$R^0_b$                            & $ 0.2170 \pm 0.0009 $ & $  1.00 $ & $ -0.20 $ & $ -0.03 $ & $  0.01 $ & $ -0.03 $ & $  0.02 $ \\
$R^0_c$                            & $ 0.1734 \pm 0.0048 $ & $ -0.20 $ & $  1.00 $ & $  0.03 $ & $ -0.07 $ & $  0.04 $ & $ -0.04 $ \\
$A^{0,b}_{\mathrm{FB}}$            & $ 0.0984 \pm 0.0024 $ & $ -0.03 $ & $  0.03 $ & $  1.00 $ & $  0.13 $ & $  0.03 $ & $  0.02 $ \\
$A^{0,c}_{\mathrm{FB}}$            & $ 0.0741 \pm 0.0048 $ & $  0.01 $ & $ -0.07 $ & $  0.13 $ & $  1.00 $ & $  0.00 $ & $  0.07 $ \\
$A_b$                              & $ 0.900~ \pm 0.050~ $ & $ -0.03 $ & $  0.04 $ & $  0.03 $ & $  0.00 $ & $  1.00 $ & $  0.08 $ \\
$A_c$                              & $ 0.650~ \pm 0.058~ $ & $  0.02 $ & $ -0.04 $ & $  0.02 $ & $  0.07 $ & $  0.08 $ & $  1.00 $ \\ 
\end{tabular}
}
\end{table}
%

%
%
%
\begin{table}[!htbp]
{ 
\caption{Measured values of uncorrelated parameters used in our fits.}
\label{tab:otherpars}
\begin{tabular}{lc}                                               
\multicolumn{1}{c}{~}                                           &
\multicolumn{1}{c}{Measured value}                              \\   \hline
%
$A_\tau$                                       & $ 0.1410   \pm 0.0064 $ \\ 
$A_e$                                          & $ 0.1399   \pm 0.0073 $ \\ 
$A_{\mathrm{LR}}$                              & $ 0.1547   \pm 0.0032 $ \\ 
$\alpha^{-1}(\mZ)$                             & $ 128.896  \pm 0.090  $ \\ 
$\alpha_{\mathrm{s}}(\mZ)$                     & $ 0.118    \pm 0.003  $ \\ 
$\mW$ (GeV)                                    & $ 80.400   \pm 0.075  $ \\ 
%
%
$\mt$ (GeV)                                    & $ 175.6    \pm 5.5    $ \\ 
$1-\mW^2/\mZ^2$                                & $ 0.2254   \pm 0.0037 $ \\ 
%
%
\end{tabular}
}
\end{table}
%

%
\renewcommand{\arraystretch}{1.3}
\begin{table}[!htbp]
\caption{Results of the fit for $\mZ$, $\mt$, $\log_{10}(\mH/{\mathrm{GeV}})$, 
         $\alpha_{\mathrm{s}}(\mZ)$, $\alpha(\mZ)^{-1}$, and $\VTB$ (second column).
         For comparison, the third column shows the results of the fit with 
         the constraint of $\VTB \equiv 1$.
         $P$ denotes the probability of obtaining a reduced chisquare 
         greater than that from the fit.}
\label{tab:results}   
{ 
\begin{tabular}{lcc}                                          
\multicolumn{1}{c}{~}                                          &
\multicolumn{1}{c}{$\VTB$ free}                                &
\multicolumn{1}{c}{$\VTB$ fixed}                                \\ \hline 
%
$\mZ$ (GeV)                              & $  91.1866 \pm 0.0020       $ & $   91.1866 \pm 0.0020       $ \\
$\mt$ (GeV)                              & $ 174.2    \pm 5.4          $ & $  172.7    \pm 5.2          $ \\
$\log_{10}(\mH/{\mathrm{GeV}})$          & $ 2.15     ^{+0.30}_{-0.39} $ & $    2.04   ^{+0.30}_{-0.37} $ \\
$\alpha_{\mathrm{s}}(\mZ)$               & $   0.1171 \pm 0.0025       $ & $    0.1188  \pm 0.0021      $ \\
$\alpha^{-1}(\mZ)$                       & $ 128.913  \pm 0.092        $ & $  128.905   \pm 0.091       $ \\
$\VTB$                                   & $   0.77   ^{+0.18}_{-0.24} $ & $ 0.9991$ (fixed)              \\ \hline
${\widetilde{\chi}}^2_0 \equiv
 \chi^2/{\mathrm{d.o.f}}$                & $   15.1 / (19 - 6)         $ & $ 16.7 / (19 - 5)            $ \\
$P({\widetilde{\chi}}^2 > 
   {\widetilde{\chi}}^2_0)$ (\%)         & $  30                       $ & $  27                   $ \\
%
%
\end{tabular}
}
\end{table}
%
%
\renewcommand{\arraystretch}{1.3}
\begin{table}[!htbp]
\caption{Correlation coefficients from the fit for 
         $\mZ$, $\mt$, $\log_{10}(\mH/{\mathrm{GeV}})$, 
         $\alpha_{\mathrm{s}}(\mZ)$, $\alpha(\mZ)^{-1}$, and $\VTB$.}
\label{tab:correl}   
{\setlength{\tabcolsep}{0.01em}
\begin{tabular}{c|rrrrrr}                                  
%
%
                                 &                    
$\mZ$                            &                    
$\mt$                            &
$\log_{10}(\mH)$                 & 
$\alpha_{\mathrm{s}}$            &      
$\alpha^{-1}$                    &                       
$\VTB$                           \\ \hline 			
$\mZ$                            & $  1.00  $ & $   0.01 $  & $   0.04 $  & $  -0.01 $  & $   0.01 $  & $   0.01 $   \\
$\mt$                            & $  0.01  $ & $   1.00 $  & $   0.65 $  & $  -0.03 $  & $   0.16 $  & $  -0.15 $   \\
$\log_{10}(\mH)$                 & $  0.04  $ & $   0.65 $  & $   1.00 $  & $   0.04 $  & $   0.65 $  & $  -0.23 $   \\
$\alpha_{\mathrm{s}}$            & $ -0.01  $ & $  -0.03 $  & $   0.04 $  & $   1.00 $  & $   0.03 $  & $   0.52 $   \\
$\alpha^{-1}$                    & $  0.01  $ & $   0.16 $  & $   0.65 $  & $   0.03 $  & $   1.00 $  & $  -0.08 $   \\
$\VTB$  		         & $  0.01  $ & $  -0.15 $  & $  -0.23 $  & $   0.52 $  & $  -0.08 $  & $   1.00 $   \\
%
\end{tabular}
}
\end{table}
%


\end{document}